\documentclass[twocolumn,showpacs,aps,prl]{revtex4}
\usepackage{epsfig,amssymb}

\begin{document}

\title{Spin transport in coupled spinor Bose gases}
\author{J.~M. McGuirk}
\affiliation{Department of Physics \\
Simon Fraser University, Burnaby, British Columbia V5A 1S6, Canada}
\date{\today}

\begin{abstract}
We report direct measurements of spin transport in a trapped, partially condensed spinor Bose gas.  Detailed analyses of spin flux in this out-of-equilibrium quantum gas are performed by monitoring the flow of atoms in different hyperfine spin states.  The main mechanisms for motion in this system are exchange scattering and potential energy inhomogeneity, which lead to spin waves in the normal component and domain formation in the condensate.  We find a large discrepancy in domain formation time scales with those predicted by potential-driven formation, indicating strong coupling of the condensate to the normal component spin wave.
\end{abstract}

\maketitle

An atomic spinor Bose gas, comprised of multiple spin species, exhibits rich dynamics not present in systems without the added spin degree of freedom.  Additionally, a partially condensed system, with a Bose-Einstein condensate (BEC) in thermal equilibrium with an uncondensed component, adds immensely to the complexity.  Such a gas can exhibit effects commonly associated with bosonic systems, including superfluidity and macroscopic phase coherence, as well as effects more typically associated with fermionic systems, namely spin waves and domain formation.  This Bose ferromagnet system is an interesting system in which to study out-of-equilibrium quantum effects resulting from a coupling between the superfluid and normal components of the gas, particularly in light of recent interest in the coexistence of superfluidity and ferromagnetism in solid-state materials \cite{ssmaterials}.

Recent work with tunable interactions in lattice-based spinor systems demonstrated phase diffusion \cite{bloch2008} and spin squeezing \cite{squeeze}, and there has been much theoretical interest, particularly in strongly interacting spinor systems.  However, even so-called weakly interacting systems with static interactions have shown complicated and surprising behavior, and these systems continue to be of interest.  Much work has been done with spinor systems at the two limits of the temperature spectrum -- in nearly pure condensates \cite{domains,spinorBEC} and in thermal gases \cite{mcguirk2002,swtheory}, but due to the complexity of the system there have been relatively few studies of partially condensed spinor systems at finite temperature \cite{finiteTtheory, sengstock2004, lewando2003b, mcguirk2003, sengstock2005, stamper2010}.   Early experiments in finite temperature systems demonstrated condensate melting driven by normal component decoherence \cite{lewando2003b}, superfluid-normal component phase locking \cite{mcguirk2003}, and thermal decoherence in coherent spin mixing \cite{sengstock2005}.  More recently, remarkable spin textures were observed in an optically confined Bose gas at finite temperature \cite{stamper2010}.  In that work, and several others using pure condensates \cite{spinorBEC,sengstock2005}, interconversion between spin states was one of the dominant effects in the dynamics.  Although interconversion leads to rich dynamics, individual spin populations are not conserved, which limits the type of information that can be obtained about the non-equilibrium dynamics of those systems. This work differs in that spin populations are strictly conserved, allowing for direct observation of spin-state flow during the unfolding spin kinetics.

In Ref.~\cite{mcguirk2003}, condensate-normal component coupling was inferred from anomalously fast domain formation as well as phase uniformity across both components.  Here we confirm these observations with quantitative spin transport measurements that definitively rule out the usual mechanism of condensate domain formation \cite{domains,domaintheory}.  As atoms confined in a magnetic trap evolve, their spins are altered by interactions with other atoms and by the confining potential itself.  By monitoring the spatiotemporal flux of atoms in the individual spin states forming the spinor, we follow energy transport driven by the spin kinetics. In particular, we measure the spatial distributions of the spin species and derive spatially dependent velocities for each spin state from the profiles, indicating a spin flux.  The mechanism for generating this spin flux is examined by comparing the measured velocities with velocities predicted from inhomogeneous transverse spin profiles, showing a large discrepancy, with the spin wave accelerating the condensate to speeds nearing the speed of sound.  This velocity discrepancy can be reconciled through the coupling of condensate domain formation to the normal component spin waves.  The presence of the condensate can be interpreted as an increased spin stiffness in the coupled system, as described in \cite{sengstock2004}.

\begin{figure}
\leavevmode
\epsfxsize=3.375in
\epsffile{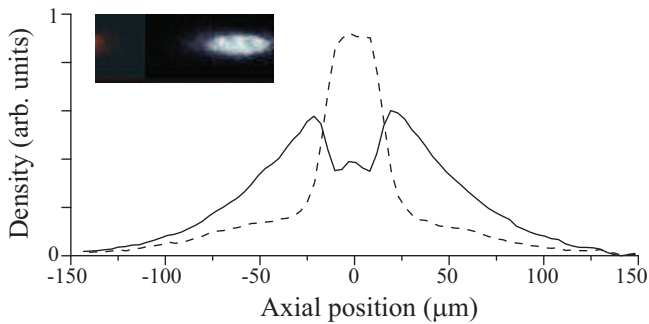}
\caption{\label{fig:density} Typical radially averaged density profiles at $T/T_c = 0.8$ for the spin states $|1\rangle$ (dashed line) and $|2\rangle$ (solid line) after a spin wave oscillation and domain formation have begun.  The $|1\rangle$ state forms domains surrounding the $|2\rangle$ state, which resides in the center of the condensate ($40~\mu$m Thomas-Fermi radius).  The inset image is a sum of the $|1\rangle$ and $|2\rangle$ images (expanded radially), showing no measurable total density fluctuation despite the spin dynamics.}
\end{figure}

The spinor system described herein is comprised of two hyperfine spin states of the $5S_{1/2}$ manifold of $^{87}$Rb atoms.  These two states, $|1\rangle \equiv |F = 1, m_{\mbox{\scriptsize{F}}} = -1\rangle$ and $|2\rangle \equiv |F = 2, m_{\mbox{\scriptsize{F}}} = 1\rangle$, form a pseudospin doublet, which may be thought of merely as a spin-1/2 system.  Bloch vector notation is used to describe the spinor \cite{allen1975}; the coordinate axes of the Bloch sphere are taken to be $u, v$ and $w$.  The ``longitudinal'' spin component $S_{\mbox{\scriptsize{w}}}$ is the population inversion, and the ``transverse'' spin components $S_{\mbox{\scriptsize{u}}}$ and $S_{\mbox{\scriptsize{v}}}$ are the coherences, with the angle in the $u-v$ plane $\phi$ being the relative phase between states $|1\rangle$ and $|2\rangle$ in a coherent superposition.  The wave-function of this system can be written $|\psi\rangle = (|1\rangle + e^{i\phi}|2\rangle)/\sqrt{2}$.

Spin kinetics in a relatively pure, trapped spinor condensate are dominated by the formation of segregated magnetic spin domains \cite{domains}.  Atom-atom interactions, characterized in the ultracold scattering limit by the $s$-wave scattering length, $a$, are responsible for domain formation.  The criterion for in-trap domain formation is that a spinor condensate's interspecies scattering lengths for the spin states must be large, positive, and not equal to the intraspecies scattering lengths \cite{domaintheory}. Domain formation does not significantly alter the overall density distribution of the condensate, merely the density distributions of the individual spin states (Fig.~\ref{fig:density} inset).

\begin{figure}
\leavevmode
\epsfxsize=3.375in
\epsffile{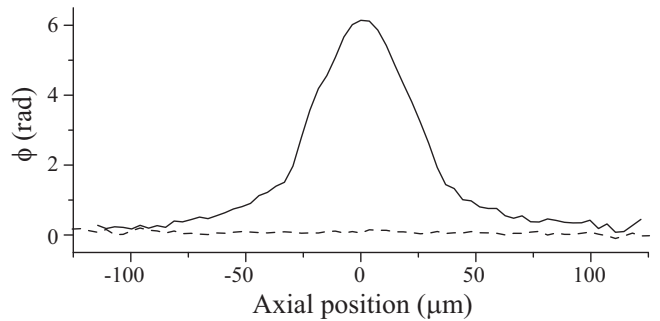}
\caption{\label{fig:phase} Radially averaged phase profiles at $t \approx 0$~ms (dashed line) and $t = 70$~ms (solid line), both at temperature $T/T_{\mbox{\scriptsize{c}}}=0.8$, obtained using Ramsey spectroscopy.  Initially, the phase is uniform across the cloud.  As time evolves, spatial inhomogeneities develop, largely due to the mean-field shift, which is dependent on the local density. The large phase bump in the center of the cloud shows the presence of the dense BEC.}
\end{figure}

The spin flux needed to form condensate domains is produced by spatial gradients in the relative phase of the spin-state superposition.  A spatial gradient in the phase of the superposition is a higher-energy configuration, and the condensate will redistribute energy by ``untwisting'' itself.  More precisely, a gradient in the condensate phase leads to a velocity, $u$, as follows \cite{dalfovo1999}:
\begin{equation}
u(\vec{r}) = (\hbar/m) \nabla \phi(\vec{r}), \label{eq:phase}
\end{equation}
where $m$ is the atomic mass.  Gradients in $\phi(\vec{r})$ arise from spatial inhomogeneities in the potential energy, primarily from inhomogeneities due to the density-dependent mean-field shift (see Fig.~\ref{fig:phase}). In the trapping geometry used here, the typical time scale for domain formation is $\sim~280$~ms for a nearly pure condensate.

The experimental setup used for this work consists of $^{87}$Rb confined in an Ioffe-Pritchard magnetic trap similar to that described in \cite{lewando2003a}.  The trap is cylindrically symmetric, with trapping frequencies of \{6.7, 236, 236\}~Hz.  Because the magnetic trap is highly anisotropic, all of the significant dynamics occur only in the loosely confined axial direction, $z$, and the density distribution may be radially averaged and effectively considered to be a one-dimensional distribution (see Fig.~\ref{fig:density}).  The temperature (and thus the fraction of atoms in the condensate) is controlled with radio frequency evaporative cooling from the critical temperature, $T_{\mbox{\scriptsize{c}}}$, down to $T/T_{\mbox{\scriptsize{c}}} < 0.3$.  The two spin states are connected by a two-photon microwave coupling \cite{mcguirk2002}.  The initial conditions for the experiments described herein are immediately following the creation of an equal coherent superposition of the $|1\rangle$ and $|2\rangle$ states via a $\pi/2$ pulse.  The initial condensate size is kept fixed at a small number, $N_0 \simeq 6\times10^4$ atoms, so as to minimize the effects of loss via dipolar relaxation in $|2\rangle$.

\begin{figure}
\leavevmode
\epsfxsize=3.375in
\epsffile{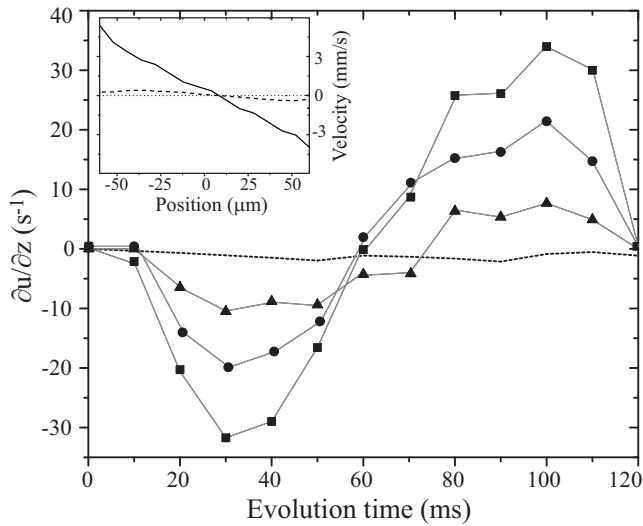}
\caption{\label{fig:velocity} Time evolution of the velocity gradients for temperatures: $T/ T_{\mbox{\scriptsize{c}}}$ = 0.8 ({\tiny $\blacksquare$}), 0.65 ({\large $\bullet$}), and 0.45 ($\blacktriangle$).  A typical predicted velocity gradient derived from the measured phase gradient is shown by the dashed line.  The inset shows a typical measured relative velocity profile at one point in time from which the gradient $\partial u/\partial z$ is derived ($T/ T_{\mbox{\scriptsize{c}}}=0.65$).  The solid line is the measured velocity, and the dashed line corresponds to the velocity predicted by the phase profile.}
\end{figure}

The spin flux is characterized by measuring the velocities of the two spin states.  This measurement is accomplished as follows.  The initial conditions are realized by creating a partially condensed condensate in a superposition of the two spin states.  Following an evolution time $t$, the spatial distribution of the population in either of the spin states can be measured (using destructive absorptive imaging).  By analyzing the time-rate of change of the density profiles of each spin state (Fig.~\ref{fig:density}), a relative velocity profile of the spin components is found.  This relative velocity, $u(z,t) \equiv u_2(z,t)-u_1(z,t)$, is derived from $\partial n_i / \partial t$ by integrating the continuity equation,
\begin{equation}\label{eq:cont}
\frac{\partial n_i}{\partial t} = -\frac{\partial}{\partial z}(u_i n_i) - \Gamma_i n_i,
\end{equation}
where $n_i(z,t)$ and $u_i(z,t)$ are the radially averaged density and velocity for state $|i\rangle$.  The final term accounts for density-dependent losses.  The inset of Fig.~\ref{fig:velocity} shows a typical velocity profile.  Because the atoms' velocities are largest near the edge of the cloud where the signal-to-noise ratio is low, a more effective characterization of the motion is through the spatial derivative of the velocity, $\partial u/\partial z$, at the center of the cloud.  The derivative contains both information about the magnitude of the velocity, as well as whether $|1\rangle$ is flowing out and $|2\rangle$ flowing in (negative slope) or vice versa.  Figure~\ref{fig:velocity} shows the time evolution of $\partial u/\partial z$ for a range of temperatures, $T/T_{\mbox{\scriptsize{c}}}$.

The first point to note in Fig.~\ref{fig:velocity} is that the velocity oscillates in time.  This oscillation is indicative of an oscillation in the spin vector, \emph{i.e.}~of spin waves \cite{swtheory,mcguirk2002}.  Spin waves exist only in the normal component, and not the condensate, due to the lack of condensate-condensate exchange scattering.  Therefore it is surprising to see signs of spin waves in the center of the ensemble, which is dominated by the presence of the condensate.  However, it was shown in \cite{mcguirk2003} that the spins of the condensate couple to the normal component spin wave through normal component-condensate exchange scattering, ``dragging'' along the spins of the condensate.

Because exchange scattering couples the condensate and normal component spins, the oscillatory nature of the velocity gradients seen in Fig.~\ref{fig:velocity} implies that condensate domain formation reverses itself as well.  Reversal of domain formation is not observed, however.  There are several reasons this may be the case.  First, the $|2\rangle$ state lifetime is approximately 150~ms for typical condensate densities due to dipolar relaxation.  This loss shrinks the $|2\rangle$ state population so that it does not exert a large effect on the $|1\rangle$ state and also has an apparent inward velocity.  Furthermore, the thermal component decoheres due to the large spatial energy inhomogeneity, as well as possibly through spin locking with the condensate, and this decoherence can melt the condensate (see \cite{lewando2003b}).  Another explanation for the lack of domain reversal is that the condensate imposes a spin stiffness to the system.  It is energetically unfavorable for the condensate spin states to interpenetrate again, and phase-separated condensate inhibits the remainder of the spin wave cycle.

The second point to note in Fig.~\ref{fig:velocity} is that the time of maximal domain formation is approximately 60~ms (when $\partial u/\partial z$ becomes zero in the center of the sample), significantly faster than domain formation time scales in pure condensates \cite{domains}.  Spin waves are further implicated as the driving mechanism in the observed spin motion by comparing the velocities derived from the measured density profiles and the potential-driven velocities predicted from the spatial phase winding.  These predictions are made in the following way.  Phase profiles such as in Fig.~\ref{fig:phase} are obtained using Ramsey spectroscopy \cite{mcguirk2002}.  A closely spaced $\pi/2-\pi/2$ pulse sequence measures the differential phases accrued across the atom cloud due to the spatially varying mean-field energy and magnetic trapping potential. These phase gradients lead to velocities, as given by Eq.~(\ref{eq:phase}).  The dashed lines in Fig.~\ref{fig:velocity} and its inset show typical predicted velocity profiles and velocity gradients derived therewith.

\begin{figure}
\leavevmode
\epsfxsize=3.375in
\epsffile{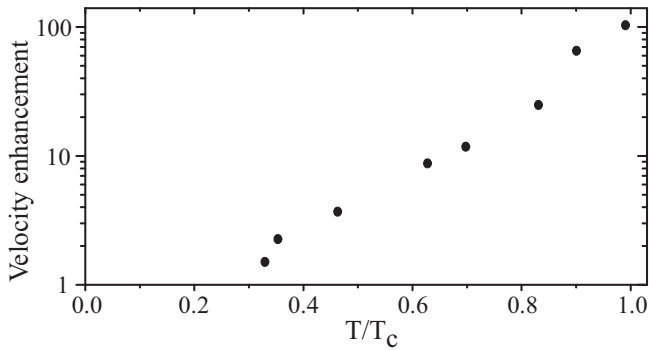}
\caption{\label{fig:velratio} Enhancement of the expected relative velocity, that is the ratio of the measured relative velocity [Eq.~(\ref{eq:cont})] to the relative velocity predicted by the spatially varying phase of the superposition [Eq.~(\ref{eq:phase})].  Values significantly greater than 1 represent a strong coupling of the condensate to normal component spin waves.  Spin waves are suppressed at low temperatures when the normal component-normal component exchange scattering rate becomes too low, and the superfluid velocity approaches that predicted by the phase profiles.}
\end{figure}

It is immediately clear that there is a large discrepancy between the measured and the predicted velocities, which leads spin domains to form much faster than expected.   In fact, the observed velocities in the condensate approach the Bogoliubov speed of sound, $v_s = \sqrt{gn/m}$, where $g = 4\pi \hbar^2 a/m$ \cite{andrews1997}.  The nature of the anomalously fast domain formation is compellingly shown by comparing the measured and predicted velocities (Fig.~\ref{fig:velratio}).  This comparison is parameterized by the ratio of the velocity gradients at the center of the cloud at the time of the peak measured velocity gradient.  For nearly pure BEC's at the lowest $T/T_c$, the velocity derived from the phase profile is in reasonably agreement with the actual measured spin velocities, but for higher temperatures the measured velocity is significantly higher than the potential-driven velocity.  This difference is even more pronounced near $T_c$, when the measured velocities become many times greater than the potential-driven motion.  Lower temperature systems agree more closely with the value expected from the phase gradients, showing a lessened effect of condensate spin exchange with the smaller normal component.  The interpretation of the velocity enhancement shown in Fig.~\ref{fig:velratio} is the entrainment of condensate domain formation by spin waves in the normal component.  Spin waves accelerate the condensate to velocities up to several orders of magnitude larger than those driven by inhomogeneities in the potential, and condensate-normal component exchange scattering couples the condensate spin to normal component spin waves.

These results agree qualitatively with the Ginzberg-Landau coupled-fluid theory presented in \cite{sengstock2004}.  At $T_c$ they predict a sharp increase in the spin stiffness, quantified by the coupled-fluid stiffness parameter $c_s$, which governs the spin wave dispersion relation $\hbar \omega = c_s k^2$ for spin wave frequency $\omega$ and wavevector $k$.  Below $T_c$, $c_s$ is predicted to be a sum of the normal component spin stiffness $c_0$ and a term roughly proportional to the condensate order parameter.  In the strong coupling limit, where the axial trap frequency is much smaller than the exchange collision frequency, $\omega_{\mbox{\scriptsize{exch}}} =gn/\hbar$, $c_0$ is given by the exchange energy according to \cite{swtheory}
\begin{equation}
c_0 = \frac{\hbar k_{\mbox{\scriptsize{b}}}T }{m w_{\mbox{\scriptsize{exch}}}}.
\end{equation}
 The explicit temperature dependence of $c_0$ is canceled by that of $k$ (proportional to the size of the atom cloud), and a constant spin wave frequency is observed as a function of temperature for fixed density \cite{mcguirk2002}.  However, as soon as there is a significant condensate component present, the spin wave frequency increases several fold.  It is difficult to make quantitative comparisons with theory below $T_c$ for several reasons.  First, as noted previously, the condensate inhibits spin wave oscillations after a half cycle, making it difficult to determine the oscillation frequency. More importantly, the observed spin waves are strongly driven in the nonlinear regime and exhibit amplitude dependence \cite{mcguirk2002,swtheory}.  Above $T_c$ it is possible to modify the excitation strength by changing the magnetic contribution to the inhomogeneous potential that drives the spin waves and to extrapolate the spin wave frequency to the low-amplitude limit.  Below $T_c$, however, the presence of the large inhomogeneous mean-field potential of the condensate makes it impossible to flatten the curvature of the driving field through similar means.

In conclusion, we have measured the relative velocities of spin components in a partially condensed spinor system.  By comparing the measured velocities with potential-driven velocities predicted from phase winding, we rule out potential-driven superfluid flow as a mechanism for domain formation in the condensate and find coupling between the condensate and normal component spin waves to be the driving mechanism for energy transport.  These results agree qualitatively with the theoretical predictions.  Additionally, while spin wave behavior drives the condensate, the spin stiffness of the phase-separated condensate inhibits further evolution of spin waves.  This complex interdependent behavior highlights the richness of a partially condensed spinor Bose gas.

The author thanks Ryan Thomas for experimental assistance.  This work was supported by NSERC and CFI.


\begin{thebibliography}{Hi!}

\bibitem{ssmaterials}
S.~S. Saxena {\it et al.}, Nature (London) {\bf 406}, 587 (2000); C.~Pfleiderer {\it et al.}, Nature (London) {\bf 412}, 58 (2001); C.~Pfleiderer, Rev. Mod. Phys. {\bf 81}, 1551 (2009).

\bibitem{bloch2008}
A. Widera {\it et al.}, Phys. Rev. Lett. {\bf 100}, 140401 (2008)

\bibitem{squeeze}
M.~F. Riedel  {\it et al.}, Nature (London) {\bf 464}, 1170 (2010); C. Gross, {\it ibid.} {\bf 404}, 1165 (2010).

\bibitem{domains}
D.~S. Hall, \emph{et al.}, Phys. Rev. Lett. {\bf 81}, 1539 (1998); J. Stenger, \emph{et al.}, Nature (London) {\bf 396}, 345 (1998).

\bibitem{spinorBEC}
See for instance T. Kuwamoto {\it et al.}, Phys. Rev. A {\bf 69}, 063604 (2004); M.-S. Chang {\it et al.}, Nat. Phys. {\bf 1}, 111 (2005); L.~E. Sadler {\it et al.}, Nature (London) {\bf 443}, 312 (2006); K.~M. Mertes {\it et al.}, Phys. Rev. Lett. {\bf 99}, 190402 (2007);  J. Kronj\"{a}ger {\it et al.}, arXiv:0904.2339 (2009); R.~P. Anderson {\it et al.}, Phys. Rev. A {\bf 80}, 023603 (2009).

\bibitem{mcguirk2002}
J.~M. McGuirk, \emph{et al.}, Phys. Rev. Lett. {\bf 89}, 090402 (2002).

\bibitem{swtheory}
M.~\"{O}. Oktel and L.~S. Levitov, Phys. Rev. Lett. {\bf 88}, 230403 (2002); J.~N. Fuchs, D.~M. Gangardt, and F. Lalo\"{e}, Phys. Rev. Lett. {\bf 88}, 230404 (2002); J.~E. Williams, T. Nikuni, and C.~W. Clark, Phys. Rev. Lett. {\bf 88}, 230405 (2002). Y. Endo and T. Nikuni, J. Low Temp. Phys. {\bf 158}, 16 (2010).

\bibitem{finiteTtheory}
H. Schmaljohann {\it et al.}, Appl. Phys. B {\bf 79}, 1001 (2004); J. Mur-Petit {\it et al.}, Phys. Rev. A {\bf 73}, 013629 (2006); A. Sinatra, Y. Castin, and E. Witkowska, Phys. Rev. A {\bf 80}, 033614 (2009) .

\bibitem{sengstock2004}
Q.~Gu, K.~Bongs, and K.~Sengstock, Phys. Rev. A {\bf 70}, 063609 (2004).

\bibitem{lewando2003b}
H.~J. Lewandowski {\it et al.}, Phys. Rev. Lett. {\bf 91}, 240404 (2003)

\bibitem{mcguirk2003}
J.~M. McGuirk {\it et al.}, Phys. Rev. Lett. {\bf 91}, 150402 (2003).

\bibitem{sengstock2005}
J. Kronj\"{a}ger {\it et al.}, Phys. Rev. A {\bf 72}, 063619 (2005).

\bibitem{stamper2010}
M. Vengalattore {\it et al.}, Phys. Rev. A {\bf 81}, 053612 (2010).

\bibitem{domaintheory}
T.~L. Ho, Phys. Rev. Lett. {\bf 81}, 742 (1998); E. Timmermans, {\it ibid.} {\bf 81}, 5718 (1998).

\bibitem{allen1975}
See for instance L.~C. Allen and J.~H. Eberly, \emph{Optical Resonance and Two Level Atoms} (Dover, New York, 1956).

\bibitem{dalfovo1999}
F. Dalfovo {\it et al.}, Rev. Mod. Phys. {\bf 71}, 463 (1999).

\bibitem{lewando2003a}
H.~J. Lewandowski, Ph.~D. thesis, University of Colorado (2003).

\bibitem{andrews1997}
M.~R. Andrews {\it et al.}, Phys. Rev. Lett. {\bf 79}, 553 (1997).
\end{thebibliography}
\end{document}